\documentclass{vgtc}

\usepackage{times}

\usepackage{color, soul}
\usepackage[normalem]{ulem}
\usepackage{amsmath}
\usepackage{amssymb}
\usepackage{makecell, array}
\usepackage[abs]{overpic}
\usepackage{graphicx}
\usepackage{multirow}
\usepackage{tabularx}
\usepackage[shortlabels]{enumitem}
\usepackage[ruled]{algorithm2e}
\usepackage{xspace}
\usepackage{booktabs}
\usepackage{subfigure}

\usepackage{float} % for [H] placement specifier

\newcommand{\fref}[1]{Fig.~\ref{#1}}

\newcommand{\cref}[1]{Chapter~\ref{#1}}

\newcommand{\ie}{i.e.,~}

\long\def\symbolfootnote[#1]#2{\begingroup%
\def\thefootnote{\fnsymbol{footnote}}\footnote[#1]{#2}\endgroup}

\newcommand{\GSView}{Interactive Gaussian Analysis View\xspace}

\newcommand{\GSTree}{Gaussian Densification Tree View\xspace}
\newcommand{\GST}{Gaussian Densification Tree\xspace}

\newcommand{\PlotView}{Property Timeline View\xspace}

\newcommand{\LogView}{Log and Control Panel\xspace}

\newcommand{\Views}{V}
\newcommand{\VisViews}[1]{\tilde{\Views}_{#1}}
\newcommand{\ViewGaussians}[1]{\tilde{G}_{#1}}
\newcommand{\Footprint}[2]{\Omega_{#1,#2}}
\newcommand{\GLoss}[2]{\mathcal{L'}(#1,#2)}
\newcommand{\PixelCover}[2]{\omega(#1,#2)}
\newcommand{\NeedleShape}[1]{s(#1)}

\newcommand{\EdgeWeight}{\lambda}

\newcommand{\ECount}[1]{E_{\mathrm{VC}}(#1)}
\newcommand{\EBlur}[1]{E_{\mathrm{B}}(#1)}
\newcommand{\ENeedle}[1]{E_{\mathrm{N}}(#1)}

\newcommand{\PV}[1]{P(#1)}
\newcommand{\Desc}[1]{D(#1)}
\newcommand{\Life}[1]{T(#1)}
\newcommand{\Family}[1]{F_{#1}}
\newcommand{\SelLineage}[1]{\Gamma(#1)}
\newcommand{\PVTau}[1]{\tau(#1)}

\newcommand{\PVWeight}{\alpha}
\newcommand{\Zoom}{z}
\newcommand{\ZoomMin}{z_{\min}}
\newcommand{\ZoomMax}{z_{\max}}
\newcommand{\ZoomSharpness}{k}
\newcommand{\SelSet}{\mathcal{G}}

\vgtcinsertpkg

\onlineid{1081}

\vgtccategory{System or Tool}

\title{Vis4GS: A Visual Analytic Tool for 3D Gaussian Splatting Reconstruction}
\author{Kai-Yuan Lin \thanks{e-mail: kenlin09250521@gapp.nthu.edu.tw} 
\and{Aryabima Mandala Putra \thanks{e-mail: aryabimamp@gapp.nthu.edu.tw}}
\and{Jui-Chi Lee \thanks{e-mail: s113065505@m113.nthu.edu.tw}}
\and{Shih-Hsuan Hung\thanks{e-mail: hungsh@cs.nthu.edu.tw}}}
\affiliation{\scriptsize{National Tsing Hua University}}

\teaser{
  \centering
  \includegraphics[width=0.96\textwidth]{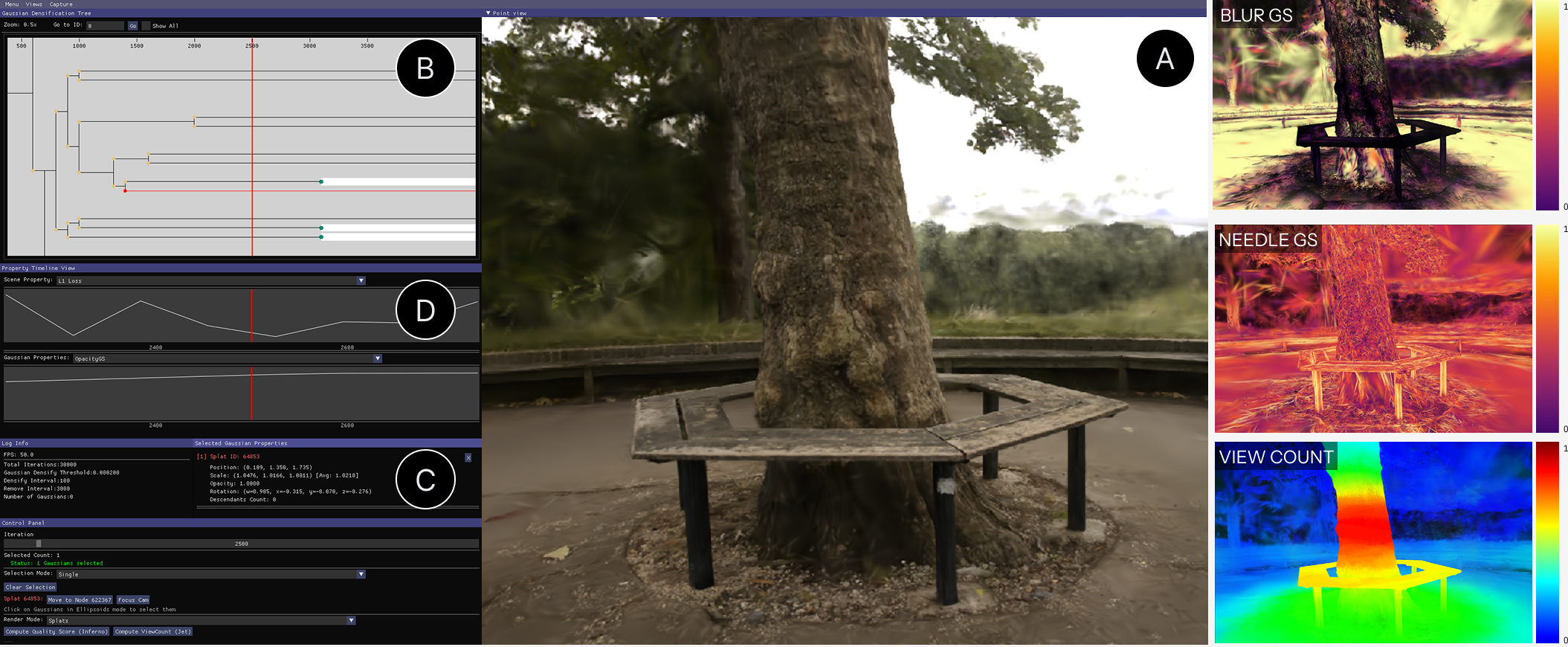}
  \caption{
  Vis4GS, our multi-view visual analytics tool for primitive-level 3DGS diagnosis. 
  The \GSView (A) combines the rendered scene with Gaussian selection and diagnostic overlays; example artifact-to-Gaussian analyses, including Blur Artifact Score, Needle-like Artifact Score, and View Coverage, are shown on the right. 
  The \GSTree (B) visualizes Gaussian genealogy and densification events, the \LogView (C) presents selected Gaussian properties and controls, and the \PlotView (D) shows training progress and selected property plots. 
  Together, these linked views support tracing visible artifacts to Gaussian properties, View Coverage, training progress, and genealogy.
  }
  \label{fig:teaser}
}
\abstract
{
3D Gaussian Splatting (3DGS) supports fast training and real-time rendering, but its optimization process remains difficult to interpret. Existing viewers mainly expose the final reconstructed scene and offer limited support for explaining how Gaussian properties contribute to visible artifacts or evolve during training. We present Vis4GS, a multi-view visual analytics tool for primitive-level diagnosis of 3DGS reconstruction artifacts. Built on the original 3DGS viewer and training framework, Vis4GS links rendered artifacts to Gaussian properties, View Coverage, training progress, and Gaussian genealogy through four linked views: an interactive Gaussian analysis view, a property timeline view, a Gaussian densification tree view, and a log and control panel. The system supports Gaussian selection, blur and needle-like artifact scoring, View Coverage analysis, and multiscale genealogy exploration of clone, split, prune, and clone-split events. By connecting scene-level artifacts with primitive-level evidence and optimization history, Vis4GS enables a structured workflow for diagnosing reconstruction failures beyond final-image inspection and global metrics. A user study also shows that Vis4GS provides stronger support for usability and artifact understanding than the original 3DGS viewer.

}

\begin{document}
\maketitle

\section{Introduction}
\label{sec:intro}

3D Gaussian Splatting (3DGS) represents a scene with explicit anisotropic Gaussian primitives initialized from sparse Structure-from-Motion points and optimized through interleaved density control and visibility-aware rendering \cite{Kerbl2023}. 
Although 3DGS supports fast training and real-time rendering, its optimization process remains difficult to interpret. 
Existing platforms provide scene inspection, editing, and basic statistics \cite{SuperSplatWeb2026,LichtFeld2026}, but offer limited support to explain how Gaussian properties produce visible artifacts or evolve during training. 
We introduce a visual analytics approach that links rendered artifacts to Gaussian properties, View Coverage, and genealogy for primitive-level diagnosis.

In a preliminary study with the original 3DGS viewer \cite{Kerbl2023}, augmented with an iteration slider and training logs, participants could detect artifacts but struggled to identify the responsible Gaussians or their evolution.
Causes such as insufficient View Coverage, poor Structure-from-Motion initialization, or densification behavior were difficult to distinguish, leading to trial-and-error diagnosis.
These observations motivate a visual analytics system that explicitly connects rendered artifacts to Gaussian primitives, quantitative cues, Gaussian properties, View Coverage, and genealogy.

From these observations, we derive the following design requirements. The system should:
\begin{enumerate}[wide=0pt, labelsep=0.5em, font=\bfseries]
    \item[\textbf{(R1)}] \textbf{Artifact-to-Gaussian attribution.}
    Trace visible artifacts in the rendered scene to the Gaussians responsible for them.

    \item[\textbf{(R2)}] \textbf{Multi-view quantitative diagnosis.}
    Provide scores, timeline plots, and linked views for inspecting Gaussian properties and View Coverage, guiding users toward suspicious regions, iterations, and Gaussians.

    \item[\textbf{(R3)}] \textbf{History-aware genealogy analysis.}
    Expose Gaussian evolution across initialization, densification, cloning, splitting, and pruning.
\end{enumerate}

To address these requirements, we propose Vis4GS, a multi-view visual analytics tool for primitive-level 3DGS diagnosis. 
As shown in \fref{fig:teaser}, Vis4GS integrates four linked components: the \GSView for artifact inspection, Gaussian selection, property analysis, artifact scoring, and View Coverage visualization (R1, R2); the \PlotView for tracking training progress and selected Gaussian properties over iterations, including PSNR, SSIM, LPIPS, and Gaussian count (R2); the \GSTree for inspecting Gaussian genealogy and densification events, including clone, split, prune, and clone-split (R3, R1); and the \LogView for accessing training logs, render controls, and selected Gaussian properties (R2). 
Together, these views support a multi-view diagnosis workflow that links rendered artifacts to Gaussian properties, View Coverage, training progress, and genealogy.
We also conducted a user study that suggested Vis4GS provides stronger support for usability and artifact understanding.
\section{Related Work}
\label{sec:related_works}
%
%Vis4GS aims to improve existing Visual Analytics using styles from different fields to achieve a more in-depth analysis for 3DGS.
%

\noindent
\textbf{Visual Analytics of 3DGS.}
Visual analytics for neural scene representations predates 3DGS; systems such as NeRVis~\cite{maas2025nervis} established precedents for interactive inspection. Existing 3DGS tools mainly support scene visualization and manipulation. SuperSplat~\cite{SuperSplatWeb2026} provides browser-based editing and Gaussian-attribute summaries, LichtFeld Studio~\cite{LichtFeld2026} integrates training, live inspection, editing, and measurement, and Splatviz~\cite{barthel2024gaussian} offers similar interactive visualization capabilities.

\noindent
\textbf{Gaussian Importance and Artifact Analysis.}
Recent 3DGS research has introduced per-Gaussian importance measures for densification, compression, and pruning. Taming3DGS~\cite{Mallick2024Taming3DGS} uses score-based sampling for budget-constrained densification, while LightGaussian~\cite{Fan2023LightGaussian}, FlexGS~\cite{FlexGS2025}, and PUP 3D-GS~\cite{Hanson2025PUP3DGS} rank primitives by significance, importance, or sensitivity. Complementary studies examine characteristic reconstruction artifacts, including blur or oversplat artifacts~\cite{Mallick2024Taming3DGS}, needle-like structures associated with highly anisotropic Gaussians~\cite{3737916.3741421}, and floaters~\cite{wang2025lowfrequencyfirsteliminatingfloating}. Visibility-aware methods such as NVGS~\cite{zoomers2026nvgs} and VAD-GS~\cite{zhang2026vadgs} further model per-Gaussian view support. Together, these studies motivate jointly analyzing Gaussian importance, shape, reconstruction error, and visibility when identifying problematic primitives.

\noindent
\textbf{Large-Tree and Multiscale Visualization.}
%{TreeSplat\cite{shen2024treesplat} utilizes a tree structure to organize Gaussian deformations hierarchically.}
Tree visualizations naturally encode hierarchies and process histories, making them suitable for Gaussian genealogy in 3DGS. Prior work improves large-tree readability through focus-and-context methods \cite{Card2002}, multiscale aggregation \cite{Elmqvist2010}, and sparse structured encodings \cite{Beurskens2025}. Aardvark \cite{Lange2024} further integrates trees with time series and images.
%, while 3DGS evolves from sparse initialization through iterative density control.
\begin{figure}[t!]
    \centering
        \includegraphics[width={\columnwidth}]{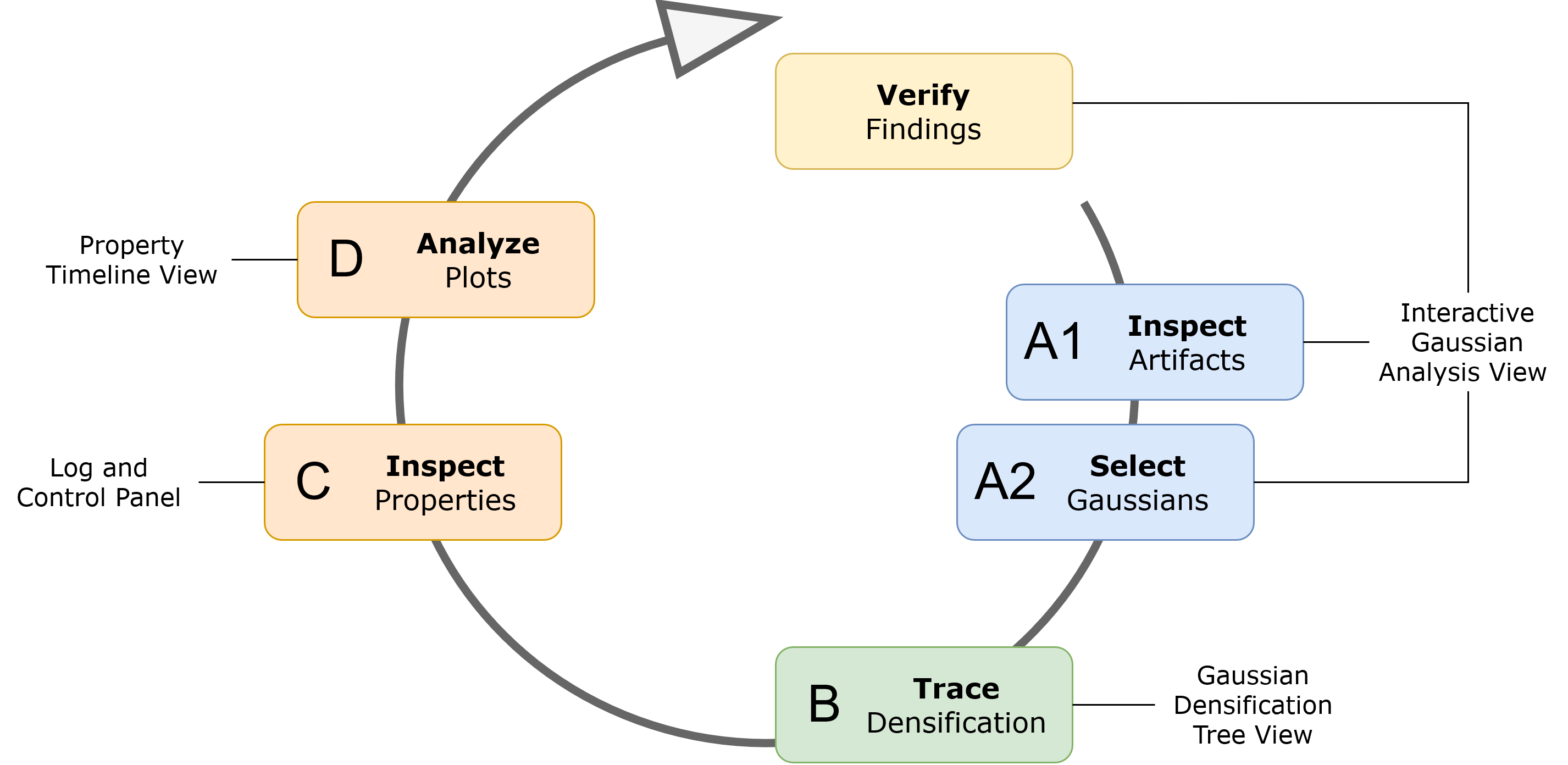}
	\caption{
User-centered diagnosis workflow of Vis4GS. 
Users inspect artifacts and select suspicious Gaussians in the \GSView (A1 and A2), trace densification history in the \GSTree (B), inspect Gaussian properties in the \LogView (C), analyze training progress in the \PlotView (D), and return to the scene to verify their findings. 
This workflow links rendered artifacts, Gaussian properties, training progress, and genealogy for primitive-level 3DGS diagnosis.
}
	\label{fig:sys}
\end{figure}

\begin{figure*}[!t]
\centering
\begin{overpic}[width=\textwidth, percent]{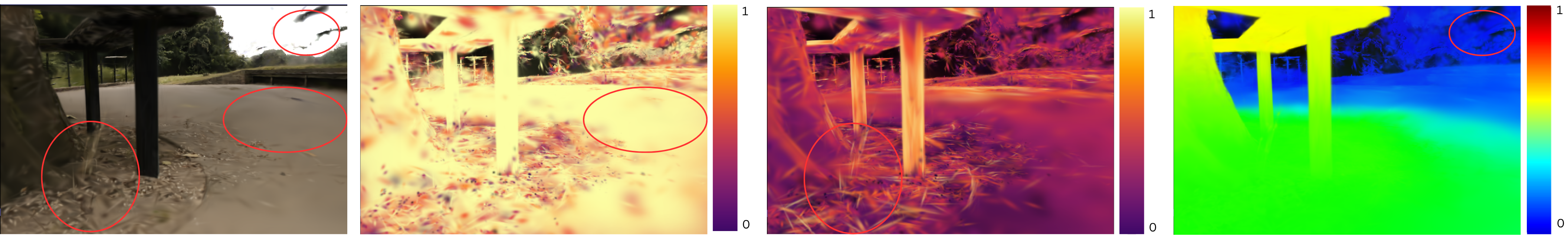}
      \put(2, -1.5){(a) 3DGS Reconstruction}
      \put(26, -1.5){(b) Blur Artifact Score}
      \put(49, -1.5){(c) Needle-like Artifact Score}
      \put(78.5, -1.5){(d) View Coverage}
    \end{overpic}
  \caption{
Artifact-to-Gaussian analysis visualizations in the \GSView. The reconstruction highlights visible artifacts (red circles), while the corresponding diagnostic scores are mapped to the scene to identify problematic primitives: (b) blur artifact score, (c) needle-like artifact score, and (d) view coverage.
}
  \label{fig:gsview}
\end{figure*}

\section{Vis4GS}
\label{sec:vis4gs}

To support primitive-level diagnosis of 3DGS reconstruction artifacts, we propose Vis4GS, a multi-view visual analytics tool built on top of the original 3DGS viewer and training framework \cite{Kerbl2023}. 
A lightweight logging module records training progress, Gaussian properties, iteration states, and densification events for downstream visualization. 
These records are visualized through four linked components that support the workflow in \fref{fig:sys}: 

\begin{enumerate}[label=(\Alph*), wide=0pt, labelsep=0.5em, font=\bfseries]
    \item \textbf{\GSView.} Users inspect rendered artifacts and select suspicious Gaussians. As the primary scene-level interface, it connects visible artifacts to the underlying Gaussian primitives through real-time rendering, property inspection, diagnostic overlays, and four selection modes: Single, Brush, Frame, and Ray. These modes support framebuffer-based picking, screen-space region selection, and octree-accelerated ray casting. Selected Gaussian IDs are synchronized across views for highlighting, property inspection, and genealogy lookup.
    \item \textbf{\GSTree.} Users trace the genealogy of selected Gaussians. It exposes clone, split, prune, and clone-split events that are hidden in standard 3DGS viewers, helping users determine whether an artifact originates from initialization or emerges during densification. The temporal genealogy shows when each Gaussian is created, cloned or split, and pruned during optimization.
    \item \textbf{\LogView.} Users configure selection, camera, iteration, and artifact-overlay settings, and inspect numerical training logs and scene properties.
    \item \textbf{\PlotView.} Users examine training metrics and temporal plots of selected-Gaussian properties, including artifact scores, opacity, and View Coverage.
\end{enumerate}

\subsection{Artifact-to-Gaussian Analysis}
\label{subsec:artifact_to_gaussian}
As shown in \fref{fig:gsview}, the \GSView (A) maps per-Gaussian diagnostic signals onto the rendered scene, helping users localize suspicious primitives before inspecting their properties or genealogy. Rather than introducing a new taxonomy, we focus on three recurring artifact patterns discussed in prior and concurrent 3DGS work: \textit{floaters}, isolated Gaussians appearing in free spac\cite{wang2025lowfrequencyfirsteliminatingfloating}; \textit{blur artifacts}, characterized by large, diffuse or oversplat-like Gaussians~\textcolor{red}{\cite{Mallick2024Taming3DGS}}; and \textit{needle-like artifacts}, associated with highly anisotropic Gaussians~\textcolor{red}{\cite{3737916.3741421}}. Accordingly, the \GSView provides three complementary per-Gaussian diagnostics: \textbf{View Coverage} for potential floaters, the \textbf{Blur Artifact Score} for large diffuse splats, and the \textbf{Needle-like Artifact Score} for elongated primitives.

View Coverage measures how broadly each Gaussian is supported by the training views and serves as a proxy for identifying underconstrained Gaussians and possible floaters. For a Gaussian $g$, we project its screen-space footprint $\Footprint{g}{v}$ into each training view $v\in\Views$ and define a \textit{visible training view} as a view in which at least one pixel of $\Footprint{g}{v}$ lies inside the valid image region after basic visibility checks, such as being in front of the camera and within the view frustum. The set of visible training views is denoted as $\VisViews{g}$, and the normalized View Coverage is defined as $\ECount{g}=|\VisViews{g}|/|\Views|$, where $\ECount{g}\in[0,1]$. A low $\ECount{g}$ indicates that Gaussian $g$ is supported by only a small fraction of training views.

Blur Artifact Score measures whether a Gaussian appears large in screen space and overlaps regions with high reconstruction and edge loss. 
We compute this score by first estimating a Gaussian-view loss and then weighting it by the Gaussian's normalized pixel coverage.
First, for each training view $v$, we define a pixel loss map by combining the L1 reconstruction error with an edge loss computed using a Laplacian filter~\cite{Marr1980Edge}:
\begin{equation}
\mathcal{L}(v)=
(1-\EdgeWeight) \mathcal{L}_1(v)
+
\EdgeWeight \mathcal{L}_{\mathrm{edge}}(v).
\end{equation}
Here, $\mathcal{L}_1(v)$ is the pixel-wise L1 error map between the rendered image and the training image, and $\mathcal{L}_{\mathrm{edge}}(v)$ is the pixel-wise difference between their Laplacian-filtered responses. 
We set $\EdgeWeight = 0.2$, following the weighting coefficient used in 3DGS~\cite{Kerbl2023}.

For a Gaussian $g$ visible in training view $v\in\VisViews{g}$, we compute its Gaussian-view loss by summing the pixel loss inside its projected footprint $\Footprint{g}{v}$, $\GLoss{g}{v}=\sum_{p\in\Footprint{g}{v}}\mathcal{L}(v)[p]$.
We also compute the normalized pixel coverage of $g$ in view $v$,
$
\PixelCover{g}{v}=
\frac{|\Footprint{g}{v}|}
{\max_{h\in\ViewGaussians{v}}|\Footprint{h}{v}|},
$
where $\ViewGaussians{v}$ denotes the set of Gaussians whose projected footprints overlap the valid image region in view $v$. 
Finally, we aggregate the Blur Artifact Score over the visible training views of $g$:
\begin{equation}
\EBlur{g}=
\frac{1}{|\VisViews{g}|}
\sum_{v\in\VisViews{g}}
\PixelCover{g}{v}\GLoss{g}{v}.
\end{equation}
A high $\EBlur{g}$ provides evidence for a possible blur or oversplat artifact.

Needle-like Artifact Score measures whether an elongated Gaussian overlaps regions with high reconstruction and edge loss. 
We first compute a shape anisotropy term from the Gaussian scale axes, and then weight it by the Gaussian-view loss defined above. 
Let the scale axes of Gaussian $g$ be $\boldsymbol{\sigma}_g=(\sigma_{g,1},\sigma_{g,2},\sigma_{g,3})$. 
We normalize the squared scales into a distribution:
\begin{equation}
q_{g,k}=
\frac{\sigma_{g,k}^{2}}
{\sum_{\ell=1}^{3}\sigma_{g,\ell}^{2}},
\quad k\in\{1,2,3\}.
\end{equation}
We compute the entropy-based effective rank ~\cite{3737916.3741421} and map it to a needle-shape score:
\begin{equation}
\NeedleShape{g} =
\frac{
3-\exp\left(-\sum_{k=1}^{3}q_{g,k}\log q_{g,k}\right)
}{2}.
\end{equation}
Here, $q_{g,k}\log q_{g,k}$ is defined as $0$ when $q_{g,k}=0$. 
The shape score $\NeedleShape{g}$ approaches $1$ when one scale axis dominates and approaches $0$ when the three scale axes are similar.

Finally, we define the Needle-like Artifact Score by aggregating the loss-weighted anisotropy over the visible training views of $g$:
\begin{equation}
\ENeedle{g}=
\frac{\NeedleShape{g}}{|\VisViews{g}|}
\sum_{v\in\VisViews{g}}
\GLoss{g}{v}.
\end{equation}
A high $\ENeedle{g}$ provides evidence for a possible needle-like boundary artifact.

%%%
\begin{figure}[b!]
    \centering
        \includegraphics[width=0.75\columnwidth]{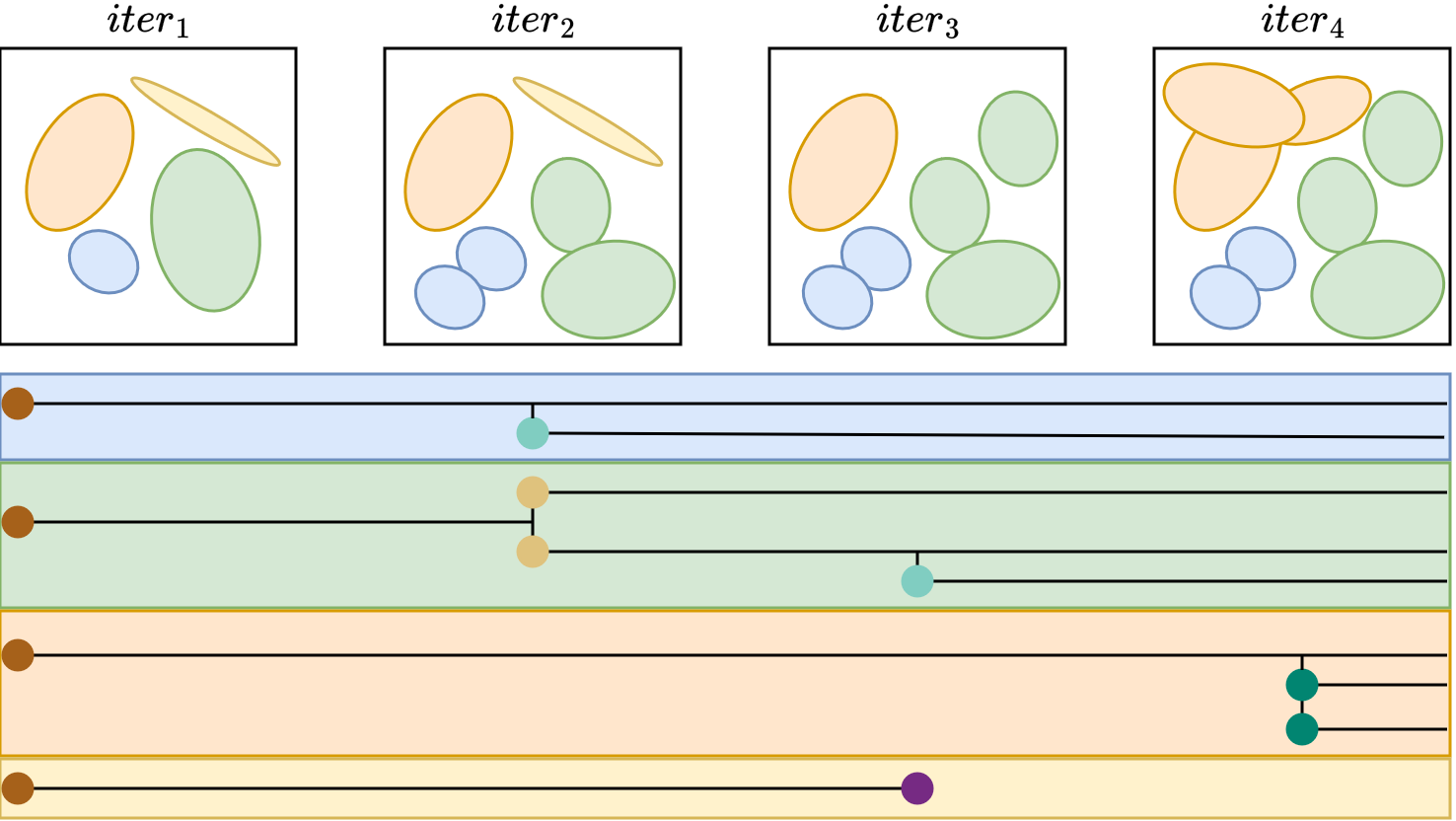}
	\caption{
Anatomy of the \GSTree using a simplified densification example. 
The top row shows four sampled iterations with creation, cloning, splitting, pruning, and clone-split events. 
The bottom row shows the corresponding tree representation, where nodes denote Gaussians, horizontal lines denote lifespans, edges denote parent-child relationships, and background bands denote Gaussian families. 
Node colors indicate root (brown), clone (aquamarine), split (tan), prune (purple), and clone-split (teal) events.
}
\label{fig:TreeAnatomy}
\end{figure}
%%%

%%%
\subsection{\GST Construction}
\label{subsec:gstree}
The \GSTree represents the densification history of 3DGS as a family-based genealogy. Each node denotes a Gaussian, each edge denotes a parent-child relationship created by densification, the horizontal position encodes the birth iteration, and a lifespan line extends to the death iteration. Node colors encode event types, including root (brown), clone (aquamarine), split (tan), prune (purple), and clone-split (teal), as illustrated in \fref{fig:TreeAnatomy}. Background bands group each root Gaussian with all of its descendants into a Gaussian family, indicate the active temporal range of that family, and terminate when the corresponding branch is pruned. 

We build the tree from exported training records by treating initial Gaussians as roots and inserting descendants according to clone, split, prune, and clone-split events. Clone events add children while keeping the parent active, split events terminate the parent and start child lifespan lines, and prune events terminate a Gaussian without descendants. Due to coarse logging intervals, clone and split events within the same interval are represented as clone-split events by inserting the observed descendants while preserving the parent until its recorded death iteration. After constructing the genealogy, we place nodes at their birth iterations and assign unique vertical positions within each family according to branch creation order, reducing overlap among nodes, lifespan lines, and connection lines while preserving readable parent-child relationships.
%%%

\begin{figure}[t]
    \centering
    \includegraphics[width=0.85\columnwidth]{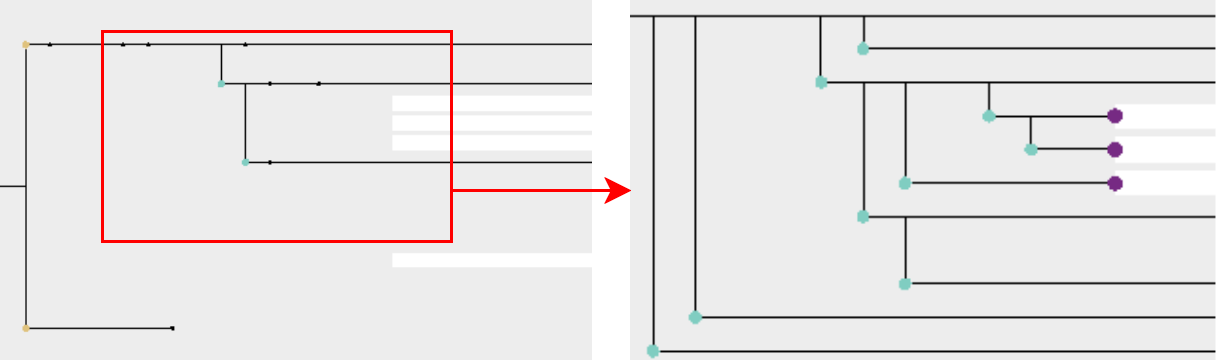}
    \caption{
    Multiscale visualization of the \GSTree. 
    The red box and arrow indicate the zoomed region; zooming out reduces clutter, while zooming in reveals local genealogy and densification details.
    }
    \label{fig:TreeMultiView}
\end{figure}

%%%

\subsection{\GST Visualization}
3DGS reconstructions may contain many Gaussian families and descendants, making the full genealogy difficult to read at all zoom levels. 
We therefore introduce a multiscale visualization that reduces clutter when users zoom out and progressively reveals local genealogy when users zoom in, as illustrated in \fref{fig:TreeMultiView}. 
Each node $i$ is assigned a perception value $\PV{i}$ based on its descendant count $\Desc{i}$ and lifespan $\Life{i}$ within its family tree, $\Family{i}$:
\begin{equation}
\PV{i}=
\PVWeight \cdot
\frac{\Desc{i}}{\max_{j\in\Family{i}}\Desc{j}}
+
(1-\PVWeight)\cdot
\frac{\Life{i}}{\max_{j\in\Family{i}}\Life{j}},
\label{eq:pV}
\end{equation}
where we set $\PVWeight = 0.4$. 
If a normalization denominator is zero, the corresponding term is set to zero.

The visibility threshold depends on the current zoom level $\Zoom$:
\begin{equation}
\PVTau{\Zoom}=
\left(
1-
\frac{\Zoom-\ZoomMin}
{\ZoomMax-\ZoomMin}
\right)^{\ZoomSharpness},
\label{eq:pv_threshold}
\end{equation}
where we heuristically set $\ZoomMin = 0.1$, $\ZoomMax = 2.0$, and $k = 1.2$.
Let $\SelSet$ denote the user-selected node set (\ie selected Gaussians), and let $\SelLineage{\SelSet}$ denote the corresponding highlighted lineage, including the selected nodes and all their ancestors and descendants. 
Nodes in $\SelLineage{\SelSet}$ remain visible regardless of the current zoom level, while all other nodes are shown only when $\PV{i}\ge\PVTau{\Zoom}$. 
This rule preserves the full selected lineage during focused analysis while filtering unrelated nodes according to the current zoom level. 
As users zoom out, the tree emphasizes long-lived or highly branching Gaussians and makes global pruning patterns easier to observe. 
As users zoom in, more local clone, split, prune, and clone-split events are revealed, enabling detailed inspection of parent-child relationships and densification history.
Tree construction dominates startup cost, while rendering remains real-time for around 4 million GS; performance and parameter details are provided in the appendix.
%%%
\section{Evaluation and Discussion}
\label{sec:evaluation}

\noindent
\textbf{User Study.}
To evaluate the usability and diagnostic value of Vis4GS, we conducted a user study with 13 participants, all of whom had prior experience with 3DGS reconstruction and the original 3DGS viewer. 
Participants completed three diagnostic tasks on assigned scenes drawn from Mip-NeRF 360 \cite{barron2022mipnerf360} and our own dataset; additional details are provided in the appendix.

After completing the tasks, participants provided System Usability Scale (SUS) ratings and 5-point Likert ratings on how well the tool supported their understanding of densification artifacts. 
Here, $\mu$ and $\sigma$ denote the empirical mean and standard deviation within each condition. 
Vis4GS achieved $\mu=74.38$ and $\sigma=9.24$, whereas the original 3DGS viewer achieved $\mu=59.58$ and $\sigma=18.06$. 
On the artifact-understanding question, Vis4GS users reported higher ratings, with $\mu=4.27$ and $\sigma=0.79$, than users of the original viewer, with $\mu=2.67$ and $\sigma=1.86$. 
Qualitative feedback was consistent with these results: participants using the original 3DGS viewer often lacked sufficient historical and Gaussian-level context, whereas Vis4GS users explicitly referred to the \GSTree, artifact score overlays in the \GSView, and the \PlotView to trace suspicious Gaussians, inspect Gaussian properties and View Coverage, and relate local artifacts to genealogy and training progress. 
Together, these findings suggest that Vis4GS provides stronger support for primitive-level artifact diagnosis.

\begin{figure}[t!]
    \centering
        \includegraphics[width={\columnwidth}]{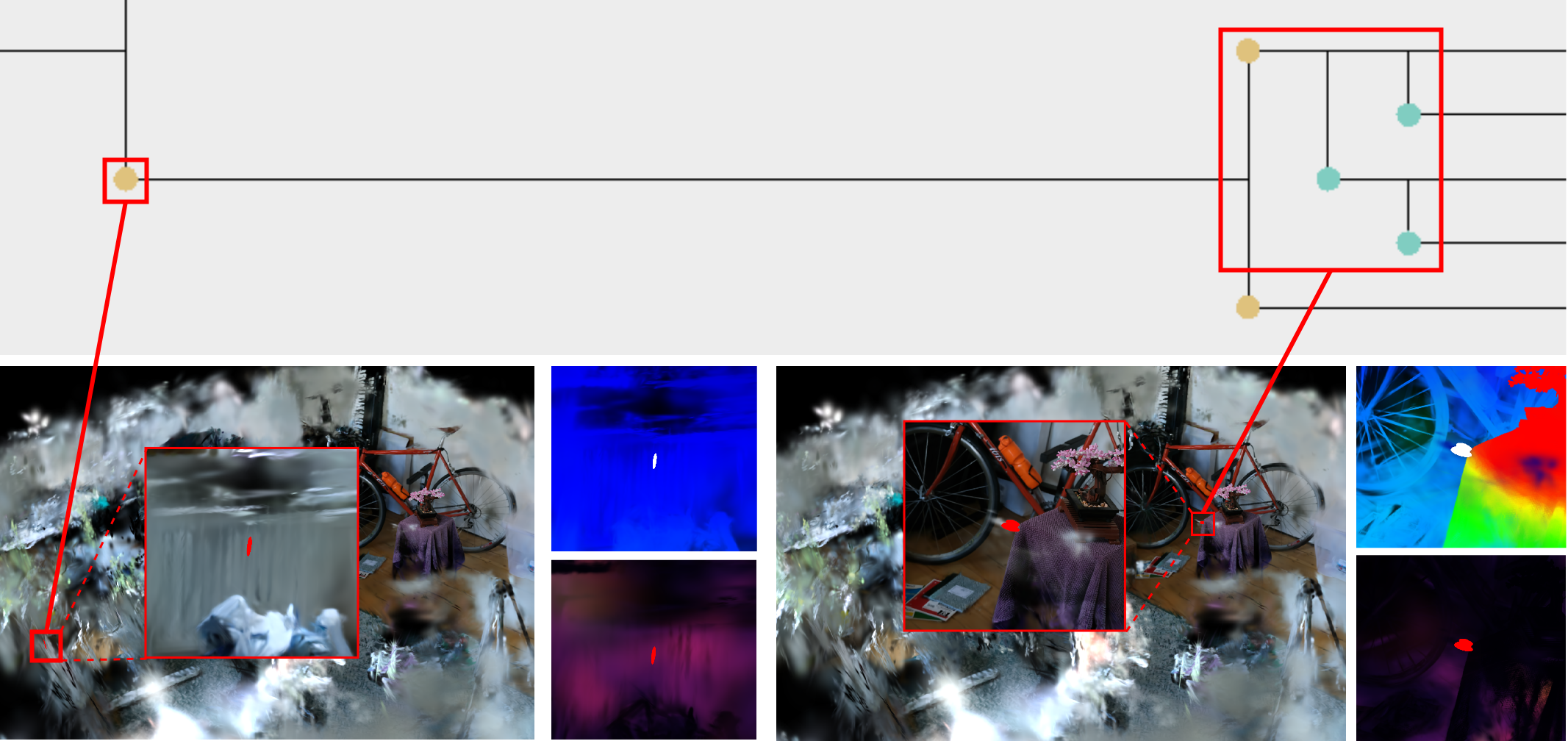}
    \caption{
    Case study of persistent floaters. 
    Linked views trace a central floater cluster to a weakly constrained ancestor Gaussian in a viewpoint-sparse region.
    }
    \label{fig:CaseStudy}
\end{figure}

\noindent
\textbf{Case Study.}
To illustrate how Vis4GS supports root-cause analysis, we examine a persistent floater artifact in \fref{fig:CaseStudy}. 
A cluster of floaters is visible near the center of the scene, but the linked views reveal that its origin is not local to that region. 
Using the \GSTree, the user can trace the artifact lineage back to an ancestor Gaussian in a peripheral area with weak View Coverage. 
The genealogy suggests that this poorly constrained Gaussian was repeatedly cloned as optimization progressed, eventually producing descendants that appear as floaters near the high-gradient center of the scene. 
This analysis shows that the artifact is not simply caused by nearby geometry, but is related to earlier densification behavior and limited training-view support. 
By linking scene inspection, View Coverage, and genealogy, Vis4GS exposes a root cause that would be difficult to identify using the original 3DGS viewer alone.

\section{Conclusion}
\label{sec:conclusion}

We presented Vis4GS, a multi-view visual analytics tool for primitive-level diagnosis of 3DGS reconstruction artifacts. 
Vis4GS links rendered artifacts to Gaussian properties, View Coverage, training progress, and Gaussian genealogy through four views: the \GSView, \PlotView, \GSTree, and \LogView. 
This workflow enables users to trace visible failures to contributing Gaussians, inspect their properties and training-view support, and relate them to densification history, making the 3DGS optimization process more interpretable beyond final-scene inspection and global reconstruction metrics. 
A user study further showed that Vis4GS achieved higher usability and artifact-understanding ratings than the original 3DGS viewer, supporting its effectiveness for primitive-level diagnosis.

Vis4GS currently focuses on diagnosis rather than correction. Although users can identify suspicious Gaussians, localized re-optimization, pruning, and re-splatting remain manual. View Coverage is only a proxy for training-view support, and the \GSTree is constrained by the temporal resolution of exported records. 
Future work will close this loop through direct corrective interventions and explore using Gaussian genealogy to guide training-time densification, pruning, and optimization, inspired by TreeSplat's tree-structured organization of Gaussian deformations~\cite{shen2026treesplat}. We will also improve scalability and modernize the rendering infrastructure.

\section*{Acknowledgements}
We thank the anonymous reviewers for their valuable comments and suggestions. This work was funded in part by the National Science and Technology Council of Taiwan (NSTC 114-2221-E-007-115-MY3).

\bibliographystyle{abbrv-doi}
\bibliography{References}

\clearpage
\appendix
\section{Tasks and Questions in User Study}
The user study involved 13 participants, primarily graduate students and researchers with experience in computer graphics, particularly 3DGS. Each session lasted approximately 30–40 minutes. The selected scenes were Bicycle and Bonsai from the Mip-NeRF 360 dataset \cite{barron2022mipnerf360} and Figurine from our own dataset.

%The user study is conducted involved 13 participants, primarily graduate students and researchers, who have experience in computer graphics, specifically 3DGS. 
%They were divided into main and control groups, where the main group (N=13) conducted the study using Vis4GS while the control group (N=6) used a base viewer for 3DGS scenes. 
%Each session lasted approximately 30-40 minutes. The selected scenes are from MiP-NeRF 360 \cite{barron2022mipnerf360} dataset and one of our own datasets: Bicycle (Outdoor), Bonsai (Indoor), and Figurine (Object-Focused). The participants are divided across the 3 scenes with 6 to 7 testers per scene. The study consists of three primary tasks:

\begin{enumerate}
    \item[\textbf{Q1.}] "A floater is already identified and selected. Using \GSView, and \GSTree, find the root gaussian responsible for this floater and report on the cause of the selected floater."

    \item[\textbf{Q2.}] "Observe this PSNR drop at iteration 9000 in the Scene Property. Using \GSView and \LogView, determine the cause of this sudden drop."
    
    \item[\textbf{Q3.}] "Explore this tool and experiment with any features provided."
\end{enumerate}

Afterwards, participants were asked to answer a system usability scale \cite{SUS} questionnaire, as well as an artifact-understanding question is \textit{"The tool helped me understand how the densification process can cause artifacts."} This is rated on a Likert scale of 1-5.

\section{System Performance and Implementation Details}
To evaluate the computational efficiency of our visual analytics tool, we benchmark performance metrics across various reconstruction scenes. All experiments are conducted on a workstation with an AMD Ryzen 9950X CPU, NVIDIA 5070 GPU, and 32 GB of RAM. Table \ref{tab:performance} summarizes the performance details of our tool across different scenes. The data shows that our tool maintains a real-time rendering and smooth navigation for the user to investigate the scene.

\begin{table}[ht]
\centering
\caption{System Performance Metrics across Scenes from Standard Datasets.}
\label{tab:performance}
\resizebox{\columnwidth}{!}{
\begin{tabular}{lccc}
\toprule
\textbf{Scene (Dataset)} & \textbf{Gaussian Count} & \textbf{Rendering Speed (FPS)} & \textbf{Loading Time (s)} \\ \midrule
Bicycle (MipNerf360) & 4.89M & 30 & 435 \\
Bonsai (MipNerf360) & 1.06M & 80 & 53 \\
Figurine (Our Dataset) & 1.07M & 85 & 58\\
\bottomrule
\end{tabular}%
}
\end{table}

\section{Parameter Explanation of \GST Visualization}
We briefly recap our equations for \GST visualization, introduced in section 3.3 of the main text. The perception value ${\PV{i}}$ of each node is defined as equation \ref{eq:pV}, and the visibility threshold that depends on the current zoom level $z$ is defined as equation \ref{eq:pv_threshold}.

\begin{equation}
\PV{i}=
\PVWeight \cdot
\frac{\Desc{i}}{\max_{j\in\Family{i}}\Desc{j}}
+
(1-\PVWeight)\cdot
\frac{\Life{i}}{\max_{j\in\Family{i}}\Life{j}}
\end{equation}

\begin{equation}
\PVTau{\Zoom}=
\left(
1-
\frac{\Zoom-\ZoomMin}
{\ZoomMax-\ZoomMin}
\right)^{\ZoomSharpness}
\end{equation}

\begin{figure*}[t]
\centering
\resizebox{\textwidth}{!}{%
\setlength{\tabcolsep}{2pt}
\renewcommand{\arraystretch}{1}
\begin{tabular}{>{\centering\arraybackslash}m{1.8cm} c c c}
& \textbf{Zoom = 0.2}
& \textbf{Zoom = 0.3}
& \textbf{Zoom = 0.4} \\[4pt]
$\alpha = 0.2$
&
\includegraphics[width=0.25\textwidth]{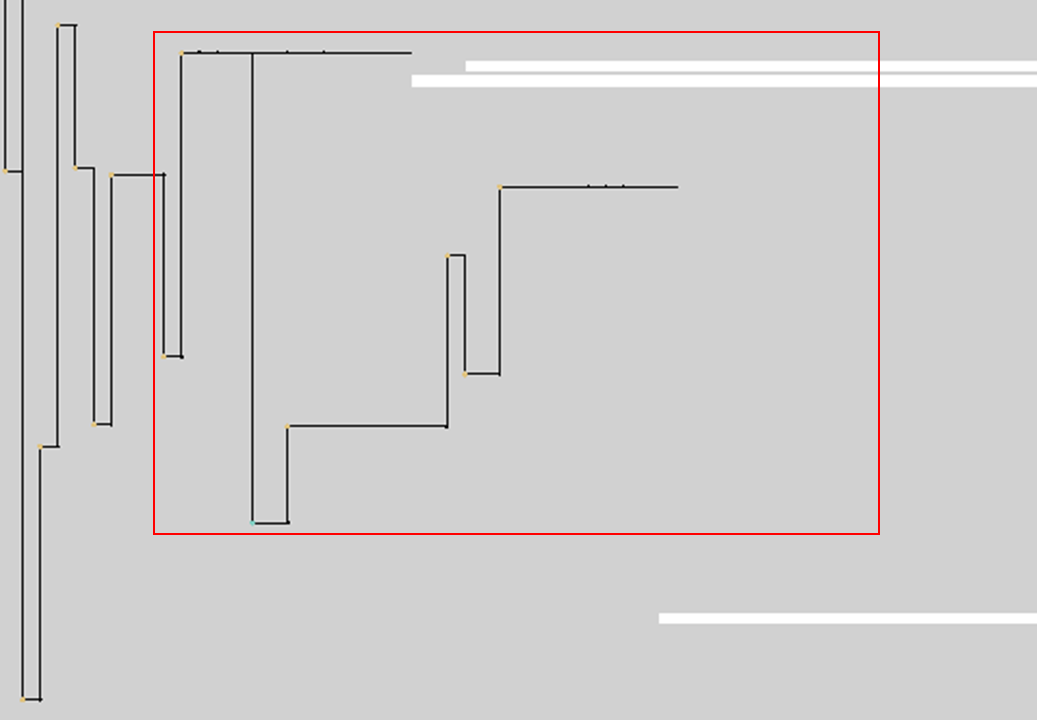}
&
\includegraphics[width=0.25\textwidth]{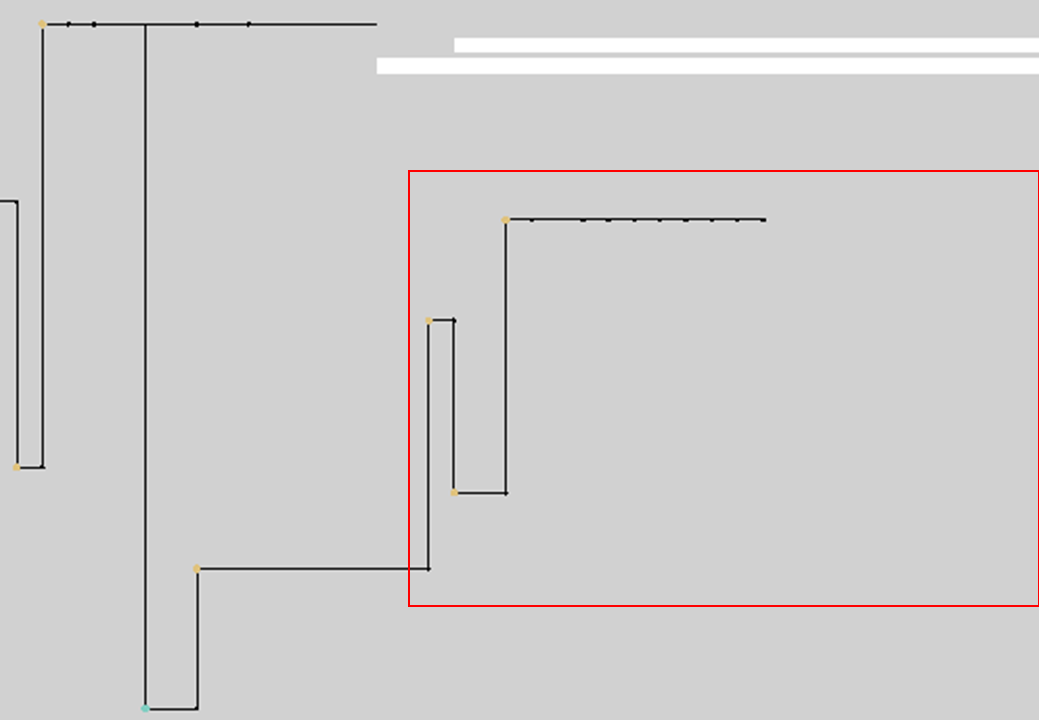}
&
\includegraphics[width=0.25\textwidth]{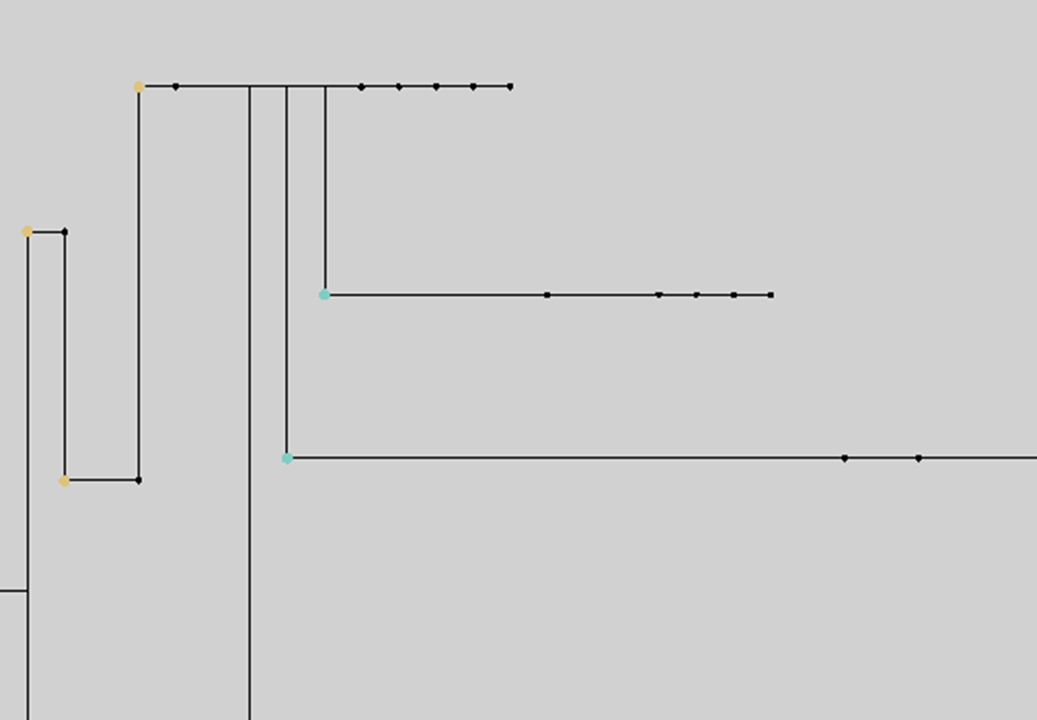}
\\[4pt]
$\alpha = 0.4$
&
\includegraphics[width=0.25\textwidth]{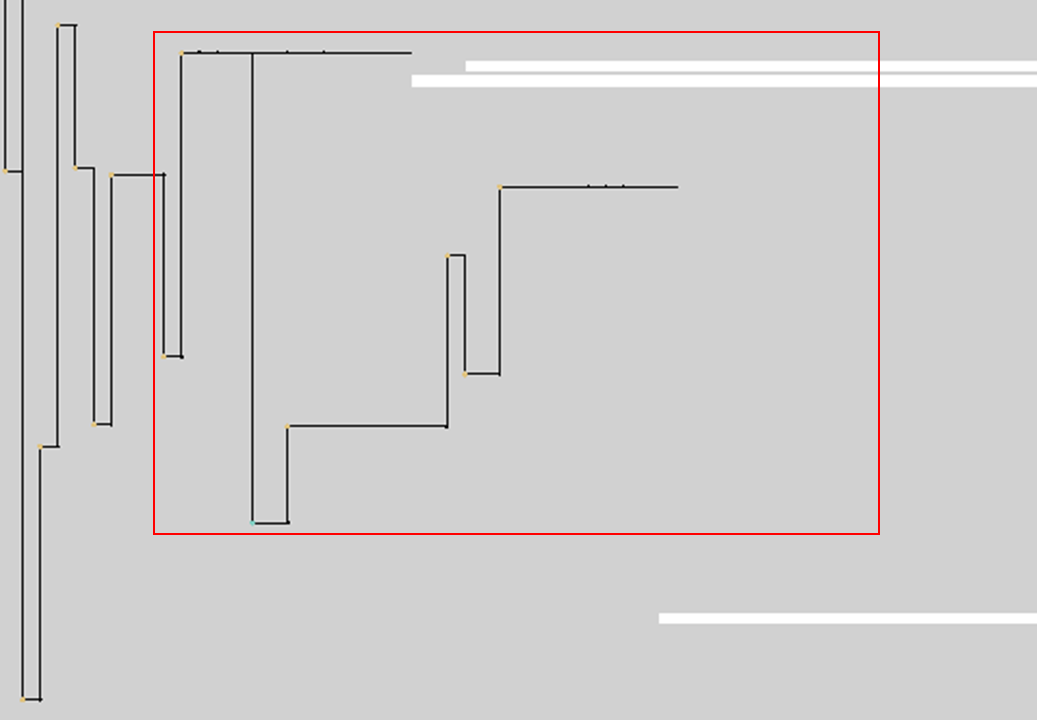}
&
\includegraphics[width=0.25\textwidth]{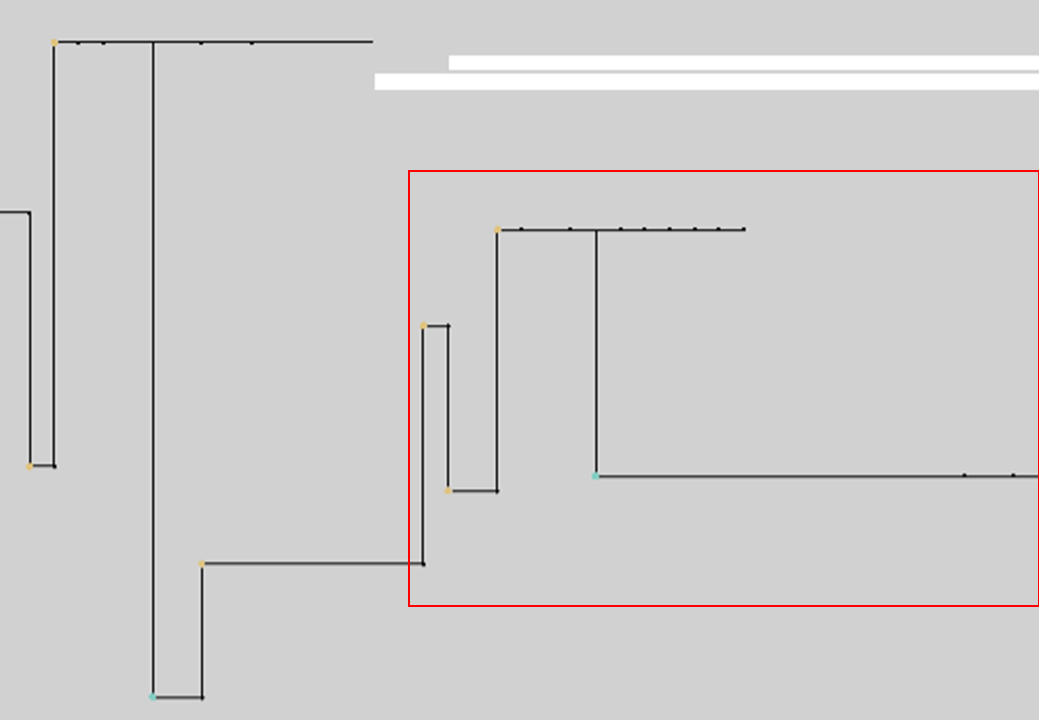}
&
\includegraphics[width=0.25\textwidth]{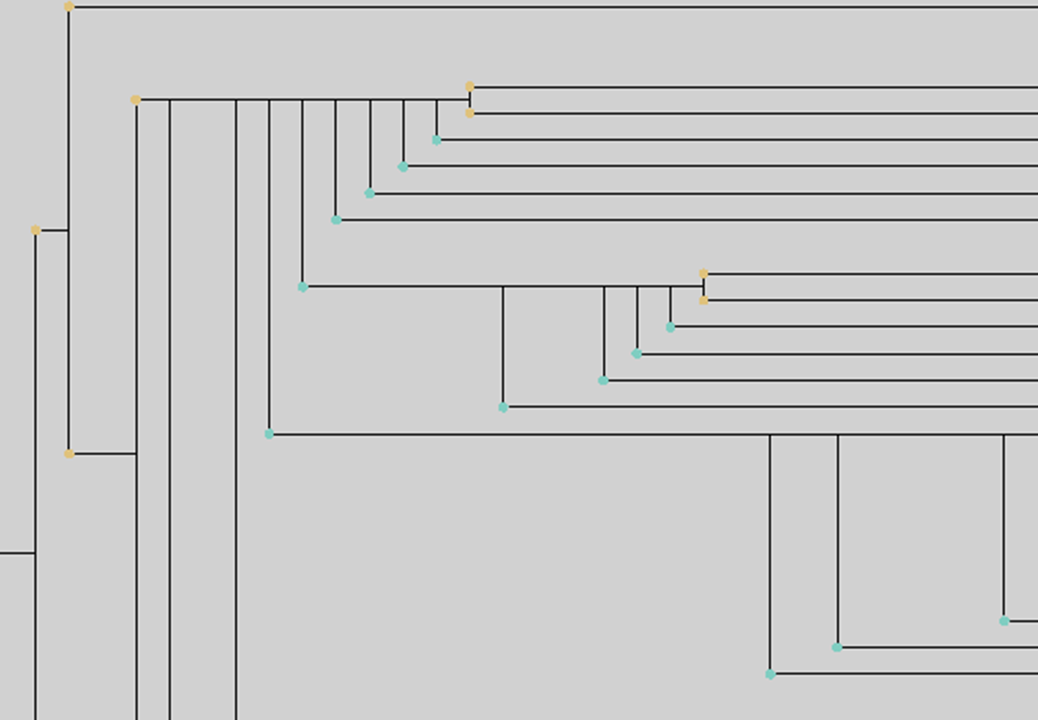}
\\[4pt]
$\alpha = 0.6$
&
\includegraphics[width=0.25\textwidth]{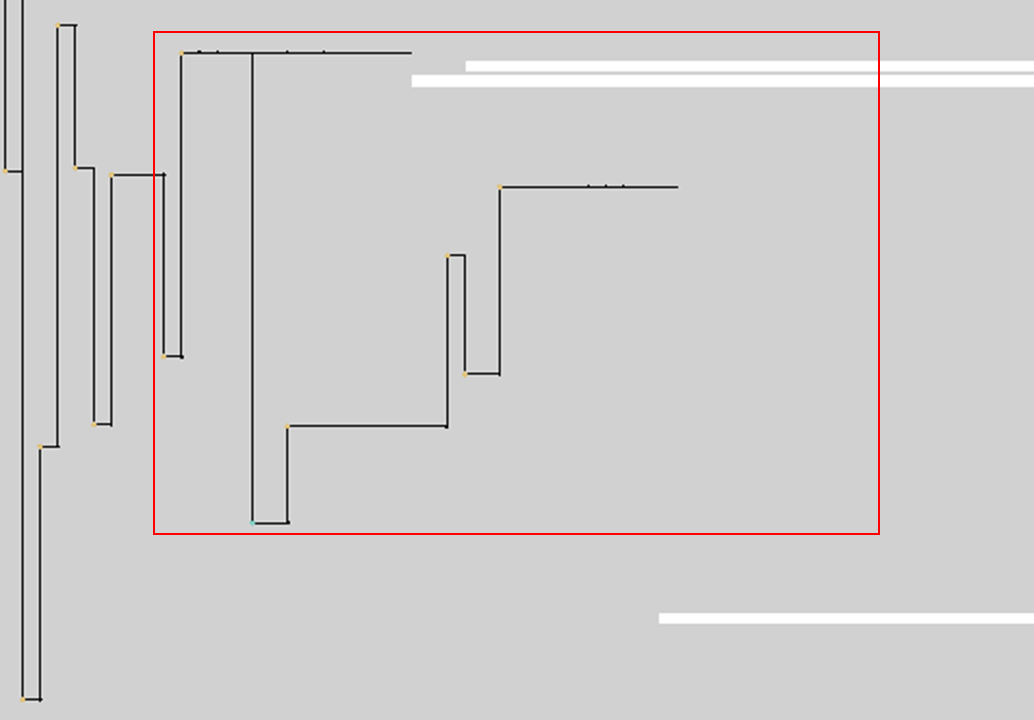}
&
\includegraphics[width=0.25\textwidth]{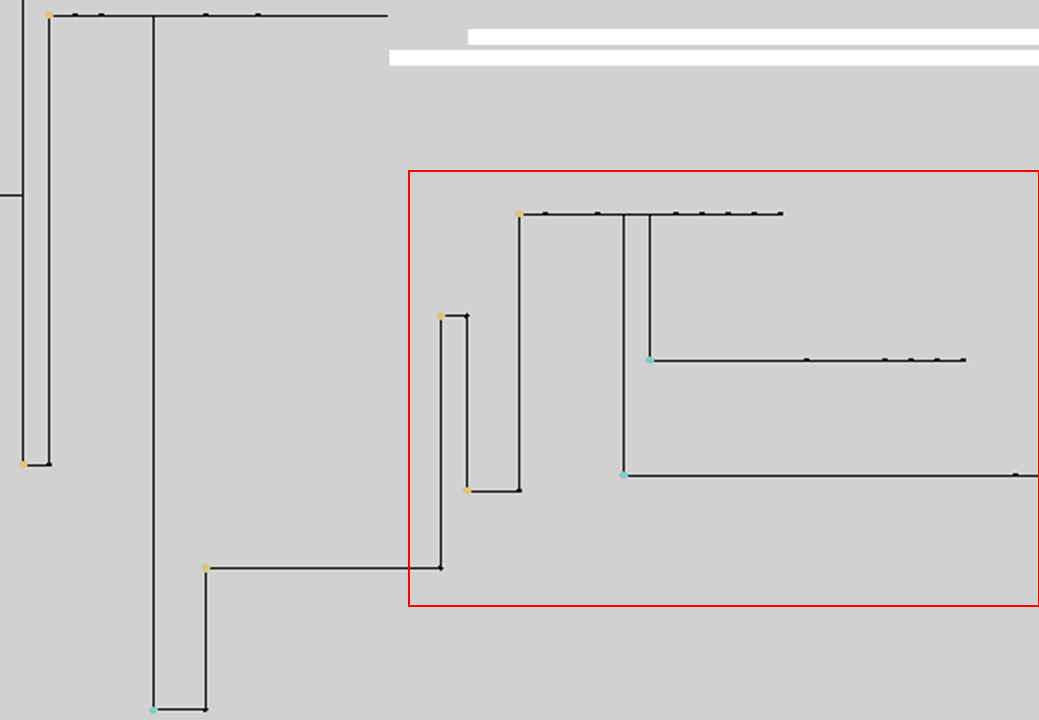}
&
\includegraphics[width=0.25\textwidth]{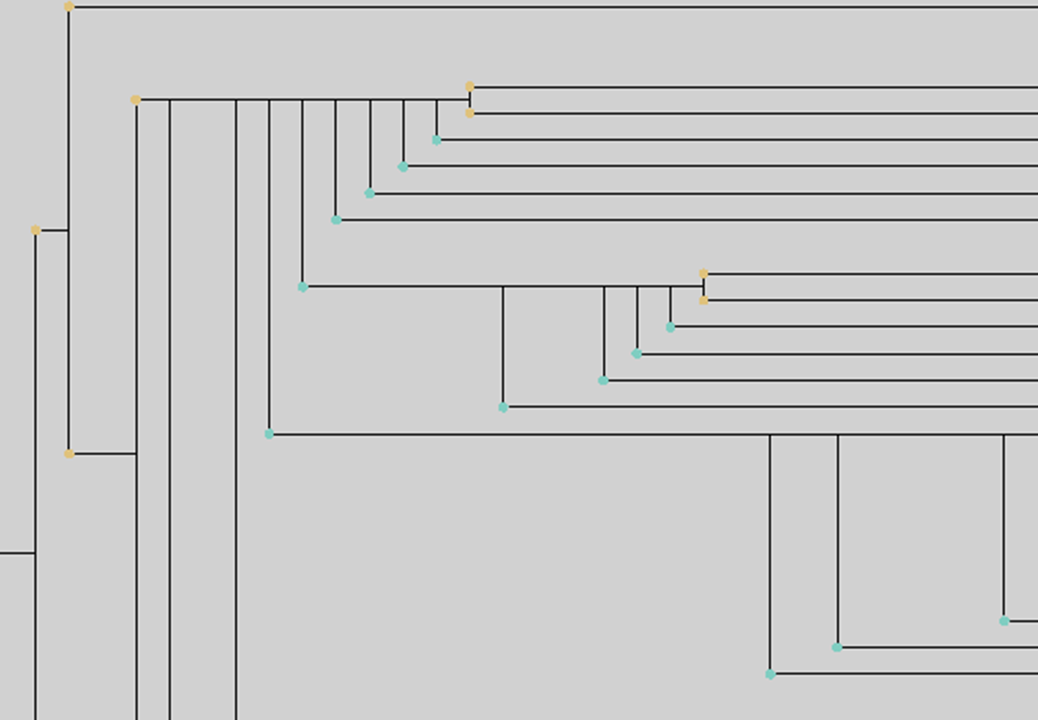}
\\[4pt]
$\alpha = 0.8$
&
\includegraphics[width=0.25\textwidth]{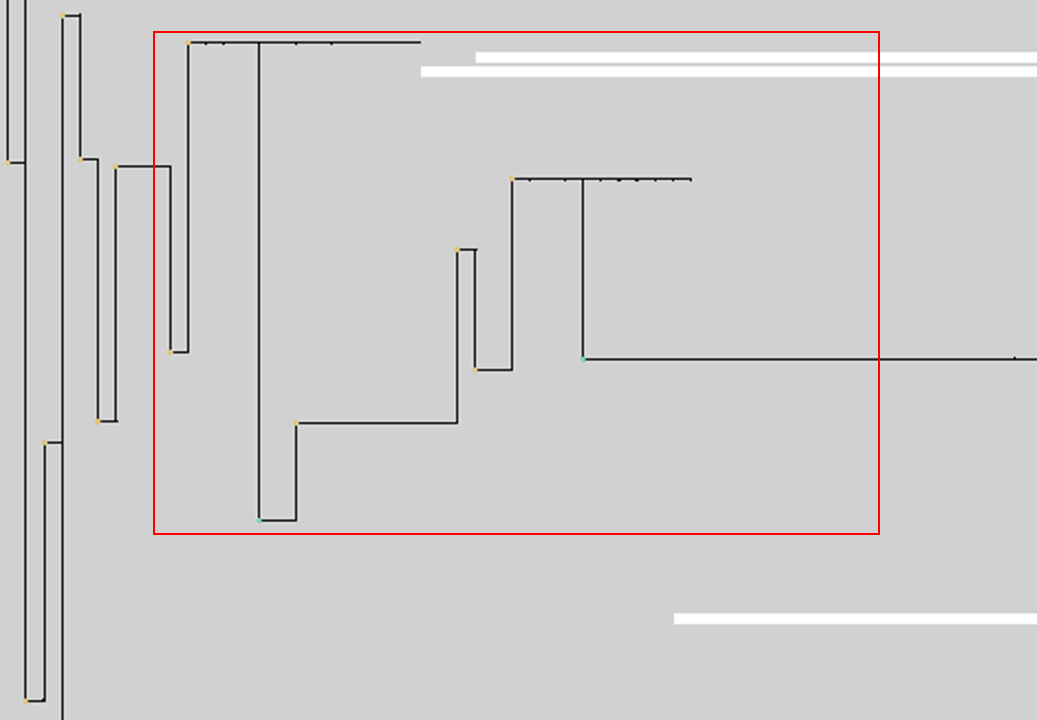}
&
\includegraphics[width=0.25\textwidth]{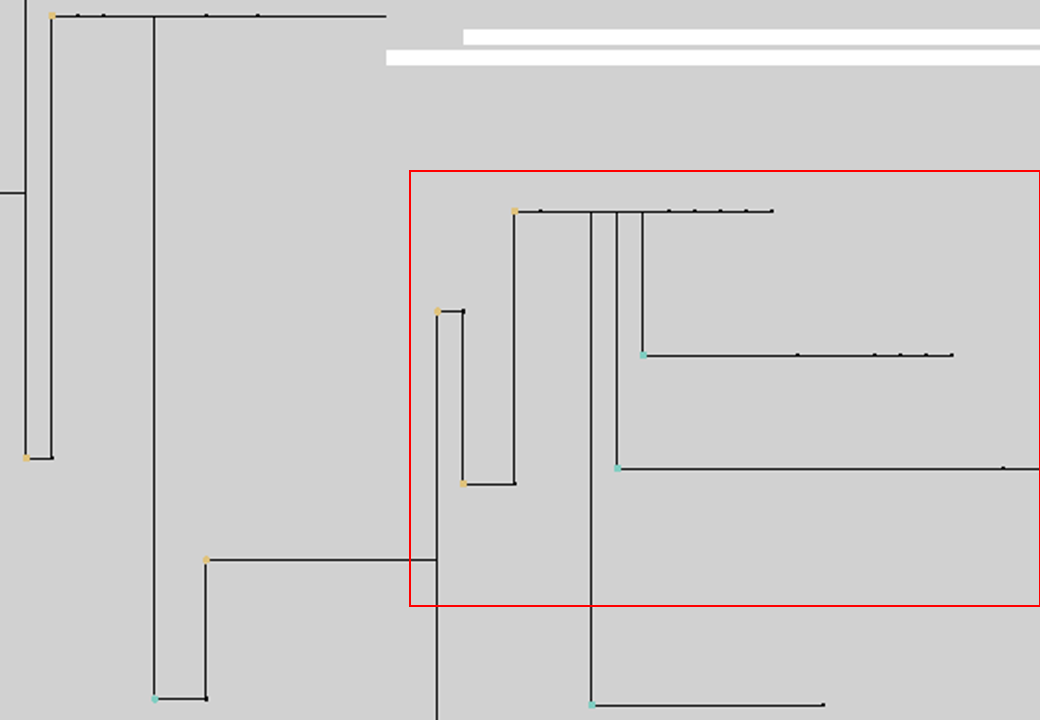}
&
\includegraphics[width=0.25\textwidth]{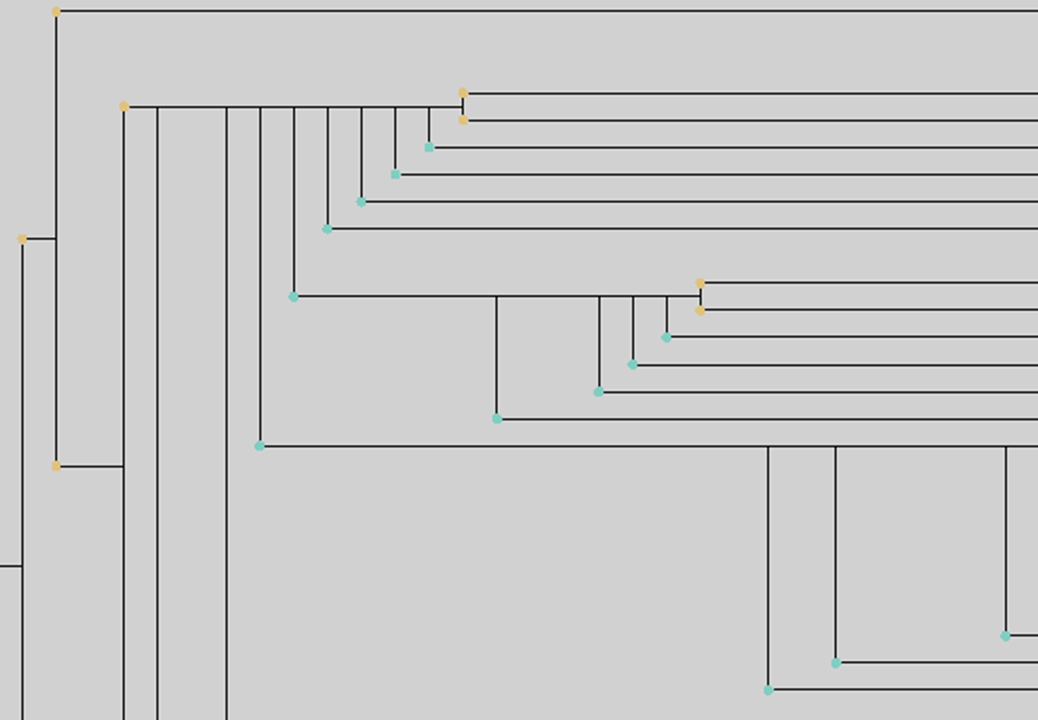}
\\
\end{tabular}%
}
\caption{
\textbf{Qualitative Level-of-Detail (LoD) tree structure sensitivity across $\alpha$ variations and zoom scales (fixed $k=1.2$).}
Rows correspond to different multi-criteria weighting factors $\alpha$, while columns show tree expansion behavior across increasing zoom levels.
}
\label{fig:lod_parameter_matrix}
\end{figure*}

We examine how the weighting factor $\alpha$ and zoom level interact to shape the Level-of-Detail tree structure. Figure~\ref{fig:lod_parameter_matrix} shows the qualitative results across four $\alpha$ values and three zoom scales with fixed $k=1.2$. Higher $\alpha$ values place greater emphasis on the descendant count $D(i)$, leading to broader tree expansions, while lower $\alpha$ values prioritize Gaussian lifetime, yielding more conservative node visibility. We set $\alpha = 0.4$ by comparing different $\alpha$ on trees of different sizes and obtained a value that shows appropriate amount of nodes at certain zoom levels. Zoom level further modulates expansion: as zoom increases, the visibility threshold $\tau(z)$ decreases, causing more nodes to become visible regardless of $\alpha$. We tested k value to obtain an acceptable smoothness of the culling from the visibility threshold.

\if 0
\section{Case Study}
To demonstrate the advantages of Vis4GS in identifying root causes of reconstruction artifacts, we demonstrate a forensic analysis of persistent floaters (see Fig. \ref{fig:CaseStudy}). A cluster of floaters can be observed near the center of the scene, which has high view-space gradients and dense camera coverage. Using our lineage-tracing feature from the \GSTree, the user discovers that these artifacts originate from an under-reconstructed Gaussian that is located in a peripheral area with sparse camera visibility. Our tool reveals a clear case of gradient-driven migration, where a poorly-constrained parent Gaussian was "pushed" toward the high-gradient center, where it underwent multiple "clone" events later on. This hierarchical insight confirms that the observed floaters are not merely local errors from nearby objects but are rather descendants from a distant, misplaced Gaussian. Without the information provided from the \GSTree, this phenomenon would not be observable to the user, which would likely lead to an ineffective adjustment to attempt to solve the issue.
%\todo{A Figure to illustrate TO BE CHECKED}

\begin{figure}[t!]
    \centering
        \includegraphics[width={\columnwidth}]{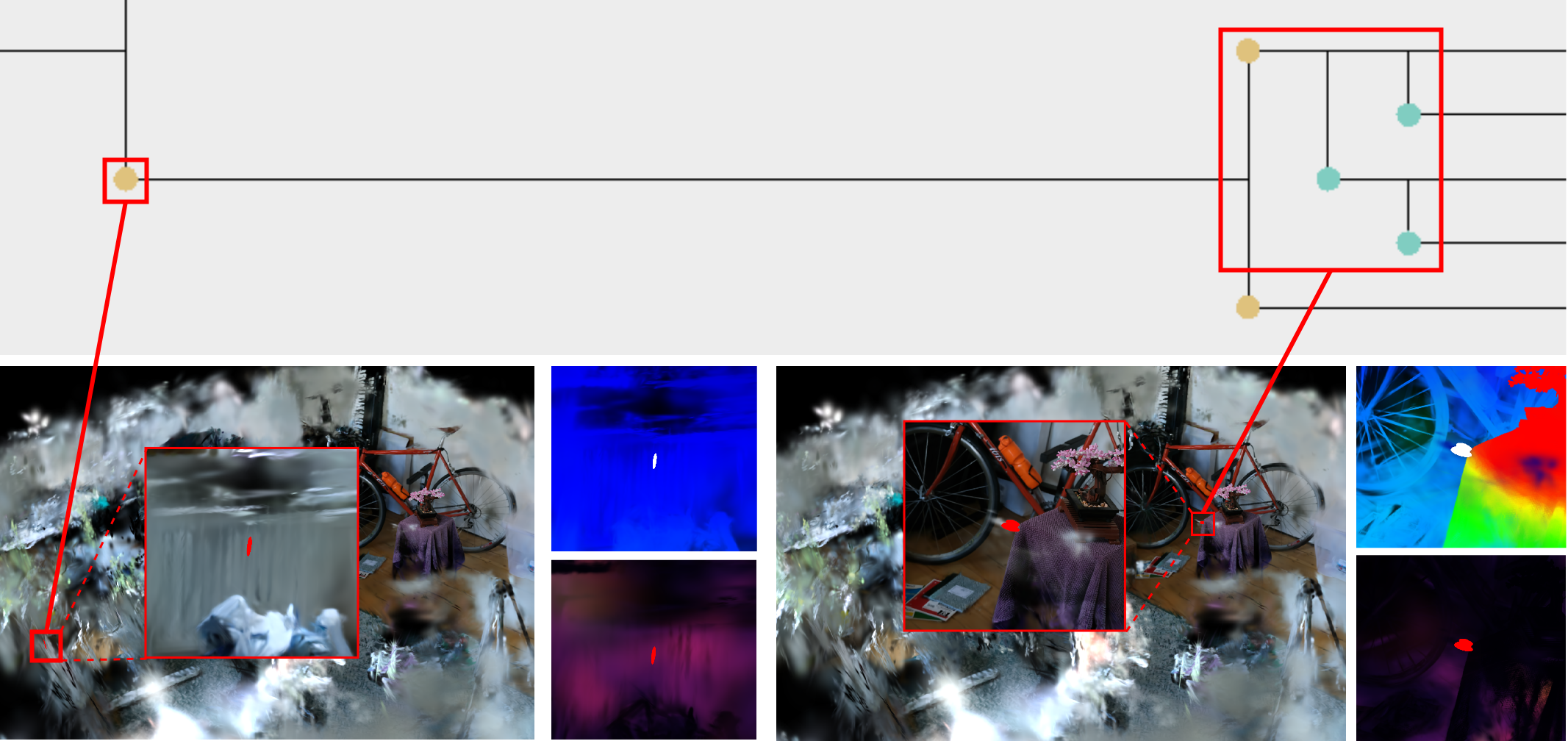}
	\caption{A central cluster of persistent floaters located near center of scene is traced to its origin, an under-reconstructed parent Gaussian in a viewpoint-sparse region of the scene, which can be seen through artifact-to-Gaussian visualization.}
	\label{fig:CaseStudy}
\end{figure}
\fi

\end{document}